\documentclass[twocolumn, pre, showpacs, english, preprintnumbers, amsmath, amssymb, superscriptaddress, aps]{revtex4-2}
\usepackage[utf8]{inputenc}
\usepackage{graphicx}
\usepackage{grffile}
\usepackage[usenames,dvipsnames]{xcolor}
\usepackage{amsmath}
\usepackage[normalem]{ulem}
\usepackage[resetlabels,labeled]{multibib}
\usepackage{appendix}
\newcites{S}{References Supplementary Materials}
\definecolor{orange}{rgb}{1,0.5,0}
\definecolor{goodgreen}{rgb}{0.1,0.5,0}
\definecolor{goodred}{rgb}{0.7,0,0}
\usepackage[pagewise]{lineno}
\newcommand{\average}[1]{\langle #1 \rangle}

\usepackage{tikz}
\usepackage{tocvsec2}
\usepackage{scrextend}
\usepackage{lineno}
\modulolinenumbers[5]
\makeatletter

\usepackage[colorlinks,urlcolor=goodgreen,citecolor=blue,linkcolor=goodred]{hyperref}

\usepackage{float}
\graphicspath{ {images/} }

\let\oldepsilon\epsilon \let\epsilon\varepsilon \let\varepsilon\oldepsilon

\makeatother

\usepackage{babel}

\newcommand{\orcid}[1]{\href{https://orcid.org/#1}{\includegraphics[width=8pt]{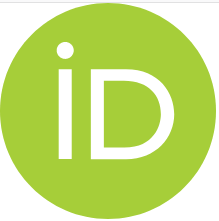}}}

\begin{document}
\title{Inductive Superconducting Quantum Interference Proximity Transistor: the L-SQUIPT}

%Authors
\author{Federico Paolucci \orcid{0000-0001-8354-4975}}
\email{federico.paolucci@sns.it}
\affiliation{NEST, Istituto Nanoscienze-CNR and Scuola Normale Superiore, I-56127 Pisa, Italy}

\author{Paolo Solinas}
\affiliation{Dipartimento di Fisica, Universit\`a di Genova, via Dodecaneso 33, I-16146,  Genova, Italy}
\affiliation{INFN - Sezione di Genova, via Dodecaneso 33, I-16146, Genova, Italy}

\author{Francesco Giazotto \orcid{0000-0002-1571-137X}}
\email{francesco.giazotto@sns.it}
\affiliation{NEST, Istituto Nanoscienze-CNR and Scuola Normale Superiore, I-56127 Pisa, Italy}

%%%%%%%%%%%%%%%%%%%%%%%%%%%%%%%%%%%%%%%%%%%%%%%%%%%%%%%%%%%%%%%%%%%%%
%% The document title should be given as usual
%% A short title can be given as a *suggestion* for running headers.
%%%%%%%%%%%%%%%%%%%%%%%%%%%%%%%%%%%%%%%%%%%%%%%%%%%%%%%%%%%%%%%%%%%%%


\begin{abstract}
The design for an \textit{inductive} superconducting quantum interference proximity transistor with enhanced performance, the L-SQUIPT, is presented and analyzed. The interferometer is based on a double-loop structure, where each ring comprises a superconductor-normal metal-superconductor mesoscopic Josephson weak-link and the read-out electrode is implemented in the form of a superconducting tunnel probe. Our design allows both to improve the coupling of the transistor to the  external magnetic field and to increase the characteristic magnetic flux transfer functions, thereby leading to an improved ultrasensitive quantum limited magnetometer. The L-SQUIPT behavior is analyzed in both the \textit{dissipative} and the \textit{dissipationless} Josephson-like operation modes,
%either in the \textit{dissipative} or in the \textit{dissipationless} Josephson-like operation mode
in the latter case by exploiting both an inductive and a dispersive readout scheme.
The improved performance makes the L-SQUIPT promising for magnetic field detection as well as for specific applications in quantum technology, where a responsive  dispersive magnetometry at milliKelvin temperatures is required.

\end{abstract}
\maketitle
\section{Introduction}

The superconducting quantum interference device (SQUID) is currently one of the most used magnetometers on the market \cite{SQUIDhand}. A SQUID consists of a superconducting ring interrupted by two Josephson junctions, thus its critical current strongly depends on the magnetic flux ($\Phi$) piercing the loop \cite{Jaklevic1964}. To achieve sizable sensitivities, SQUIDs typically employ large pickup loops, yielding a best intrinsic flux noise on the order of $\sim1$ $\mu\Phi_0/\sqrt{\text{Hz}}$ \cite{SQUIDhand}, where $\Phi_0=2.067\times10^{-15}$ Wb is the magnetic flux quantum. Differently, scanning nanoscale SQUIDs showed a flux noise as low as 50 n$\Phi_0/\sqrt{\text{Hz}}$ thanks to low inductance loops and the vicinity to the magnetic moment source \cite{Vasyukov2013}.

Last decade witnessed the advent of another sensitive magnetometer: the superconducting quantum interference proximity transistor (SQUIPT) \cite{Giazotto2010}. It is realized in the form of a superconducting ring embodying a normal metal (SNS) \cite{Meschke2011,Ronzani2014,Jabdaraghi2014,Sophie2015,Jabdaraghi2016,Jabdaraghi2017} or superconducting (SS$_1$S) \cite{Virtanen2016,Ronzani2017,Ligato2017} nanowire Josephson junction. Thanks to the superconducting proximity effect \cite{DeGennes,Buzdin2005}, a phase-dependent minigap ($E_g$) appears in the density of states (DoS) of the nanowire \cite{Zhou1998}. The latter is modulated by the superconducting phase difference built across the nanowire generated by the magnetic flux piercing the superconducting ring. In the SQUIPT, the read-out operation is typically performed by recording the $\Phi$-dependent current versus voltage characteristics of a tunnel probe (normal metal or superconductor) directly coupled to the proximitized Josephson junction. 

The best experimental sensitivity achieved so far in a SNS-SQUIPT reaches $500$ n$\Phi_0/\sqrt{\text{Hz}}$ at 240 mK \cite{Ronzani2014}, while it gets about 260 n$\Phi_0/\sqrt{\text{Hz}}$ at 1 K for a SS$_1$S device \cite{Ronzani2017}. These values are a few orders of magnitude larger than the limiting theoretical flux noise of about 1 n$\Phi_0\sqrt{\text{Hz}}$ \cite{Giazotto2011}, because the SQUIPT magnetometer shows different weaknesses and structural drawbacks. 
In particular, the SQUIPT suffers from low coupling between the external magnetic field and the superconducting loop. Indeed, a conventional device needs a small loop in order to have a negligibly small ring inductance compared to the junction Josephson inductance. Only under this assumption, the full phase bias occurs across the proximitized junction thereby allowing an efficient flux-induced modulation of the DoS of the weak-link. Furthermore, the phase biasing of the junction is efficient only for an almost-sinusoidal current-phase-relation (CPR), since in such a case the Josephson inductance of the weak-link at $\Phi_0/2$ is always finite \cite{Golubov2004,Likharev1979}. By contrast, a sizable sensitivity of the SQUIPT would be achieved by exploiting nanowire junctions in the \textit{short} limit ($\Delta_0\leq\hbar D/L^2$, where $\Delta_0$ is the zero-temperature gap of the superconductor, $\hbar$ is the reduced Planck constant, while $D$ and $L$ are the diffusion constant and the physical length of the nanowire, respectively) \cite{leSueur2008,Heikkila2002,Virtanen2016}, since in this regime the CPR is a non-sinusoidal function of the phase ($\varphi$) \cite{Golubov2004,Likharev1979}. Yet, in the \textit{short} limit and for low temperatures, the Josephson inductance is effectively vanishing at $\Phi\to\Phi_0/2$ thus preventing the full phase biasing of the junction. Therefore, a conventional SQUIPT needs to be operated at higher temperatures, where the Josephson inductance is finite, but the sensitivity can be sizeably reduced. Furthermore, fully superconducting SQUIPTs in the \textit{long}-junction limit present CPRs hysteretic with direction of the external magnetic flux \cite{Virtanen2016,ligato2020}, thus hampering their application as magnetometers. 

Here, we propose an inductive superconducting quantum interference proximity transistor (i.e., the L-SQUIPT) that solves the above described intrinsic limitations typical of conventional SQUIPT magnetometers. To this end, the L-SQUIPT takes advantage of a double loop geometry to efficiently bias the second SNS Josephson junction assumed to be in the \textit{short} limit. Furthermore, by employing a superconducting tunnel probe, the L-SQUIPT read-out operation can be realized either through a dissipative (quasiparticle tunneling) or via dissipationless (Josephson supercurrent) measurements depending on the requirements of the specific application. The L-SQUIPT is predicted to show a best quantum limited noise as low as a few n$\Phi_0/\sqrt{\text{Hz}}$, thus improving the sensitivity achievable
with conventional SQUIPT and SQUID magnetometers. This makes the  L-SQUIPT potentially relevant for magnetic field detection as well as for other applications in the field of quantum technologies \cite{polini2022materials}.

The article is organized as follows: Sec. \ref{Structure} presents the structure of the L-SQUIPT and the basic equations describing the SNS Josephson junctions embedded in the superconducting rings; Sec. \ref{Phase-Bias} shows the phase-biasing of the output SNS Josephson junction by the external magnetic flux; Sec. \ref{Dissipative} describes the dissipative read-out of the L-SQUIPT in both voltage and current bias operation; Sec. \ref{LessRead} presents the dissipationless read-out realized by means of inductive and dispersive measurement schemes; and Sec. \ref{conclus} resumes the concluding remarks.

\begin{figure} [t!]
    \begin{center}
            \includegraphics [width=\columnwidth]{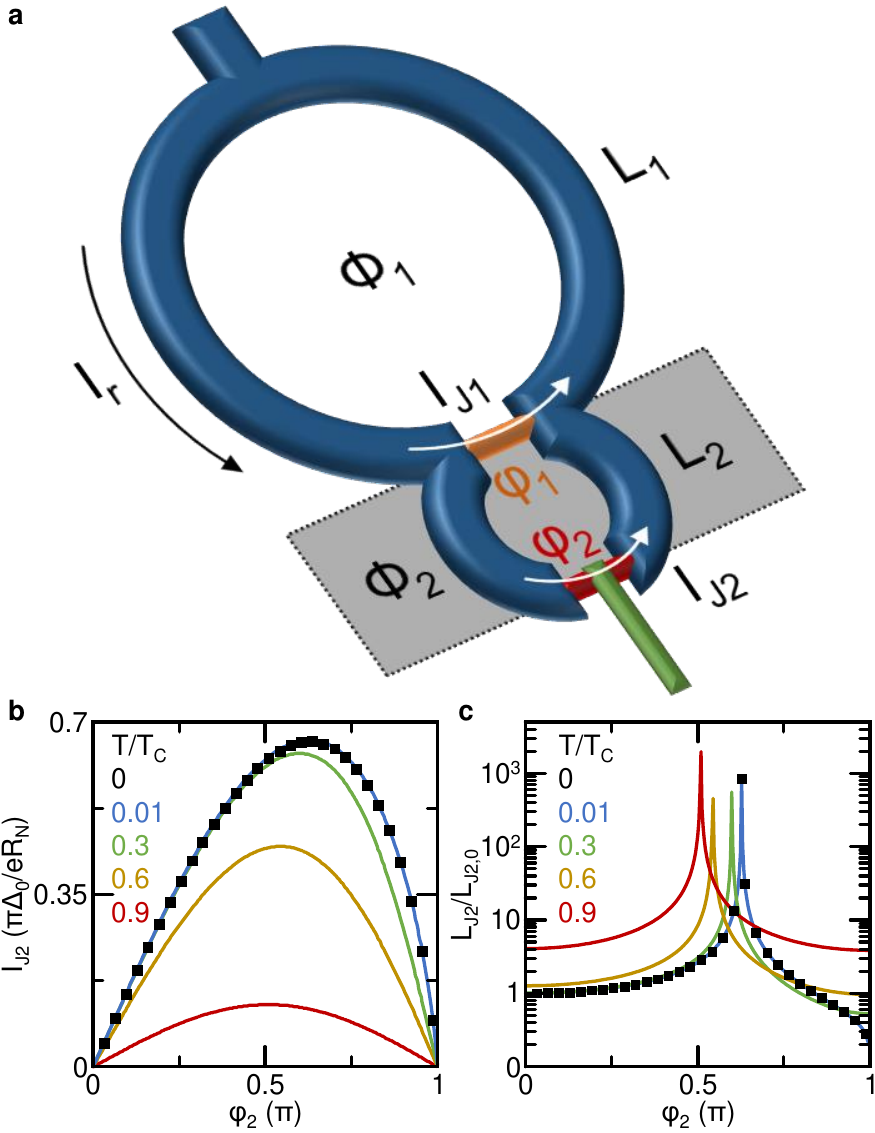}
    \end{center}
    \caption{\textbf{(a)} Scheme of the L-SQUIPT. The device is composed by two superconducting loops of inductance $L_1$ and $L_2$ interrupted by the two SNS Josephson junctions ($J_1$ and $J_2$) of critical current $I_{J1}$ and $I_{J2}$, respectively. The second ring is supposed to be screened against the external magnetic field ($\Phi_2=0$). Therefore, the phase drop across the two junctions is equal ($\varphi_1=\varphi_2$).
    \textbf{(b)} Kulik-Omel'yanchuk current-to-phase relation (KO-CPR) of $J_2$ calculated in the short junction limit for different values of temperature.  
    \textbf{(c)} Normalized Josephson kinetic inductance ($L_{J2}/L_{J2,0}$) calculated from the KO-CPR. $L_{J2,0}$ is the zero-temperature and zero-phase kinetic inductance of J$_2$.}
\label{Fig1}
\end{figure}

\section{Structure}
\label{Structure}
The L-SQUIPT is composed of two superconducting loops each of them interrupted by a normal metal weak-link forming a SNS Josephson junction, as shown in Fig. \ref{Fig1}-a. The first loop (of inductance $L_1$) converts the external magnetic flux ($\Phi_1$) into a superconducting phase drop ($\varphi_1$) across the Josephson junction ($J_1$, orange), i.e., it operates a flux-to-phase conversion ($\Phi_1\to\varphi_1$). In order to have an efficient coupling to the external magnetic field, the first superconducting loop needs, in principle, to be sufficiently large. To optimize the $\Phi_1\to\varphi_1$ conversion, the CPR of $I_{J1}$ is supposed to be sinusoidal (\textit{long}-junction limit), that is \cite{Golubov2004}
\begin{equation}
    I_{J1} (T)=I_{C1}(T) \sin{\left(\varphi_1\right)},
    \label{Icrit}
\end{equation}
where $I_{C1}(T)$ is the temperature-dependent critical current of the $J_1$ junction and $T$ is the temperature. 
A simplified equation for the critical current of $J_1$ can be found in the high temperature regime, that is for $k_B T > 5 E_{Th}$ where $k_B$ is the Boltzmann constant and $E_{th}=\hbar D/L^2$ is the Thouless energy (with $\hbar$ the reduced Planck constant, $D$ the diffusion coefficient of N and $L$ the physical length of the junction). In this limit it reads
\begin{equation}
    I_{C1}(T)=\frac{64 \pi k_B T}{eR_1}\sum_{n=0}^{\infty}\frac{\sqrt{\frac{2 \omega_n}{E_{Th}}}\Delta^2(T)\exp{\left[-\sqrt{\frac{2 \omega_n}{E_{Th}}}\right]}}{[\omega_n+\Omega_n+\sqrt{2(\Omega_n^2+\omega_n\Omega_n)}]^2},
    \label{AmbeEq}
\end{equation}
where $e$ is the electron charge, $R_1$ is the normal-state resistance of the junction, $\Delta(T)$ is the temperature-dependent superconducting energy gap of the ring, $\omega_n(T)=(2n+1)\pi k_B T$ is the Matsubara frequency, and $\Omega_n(T)=\sqrt{\Delta^2(T)+\omega_n^2(T)}$. 

The second loop (of inductance $L_2$) is supposed to be fully screened from the external magnetic field (thus $\Phi_2=0$), for instance through a superconducting plate (grey rectangle in Fig. \ref{Fig1}-a), and operates as a phase-to-phase ($\varphi_1\to\varphi_2$) transformer. 
$L_2$ needs to be sufficiently small in order to limit the phase drop along the smaller superconducting ring and, thus, to maximize the efficiency of the $\varphi_1\to\varphi_2$ transformation. The junction $J_2$ (red) is supposed to be in the \textit{short} limit, thus obeying to the Kulik-Omel’yanchuk (KO) model \cite{Kulik1975}. Therefore, the temperature dependent CPR of a $J_2$ takes the form
\begin{equation}
     I_{J_2} (\varphi_2, T)=\frac{\pi\Delta(T)}{eR_2}\Xi(\varphi_2,T),
     \label{KO}
\end{equation}
where $R_2$ is the normal-state resistance of $J_2$. The phase dependence in the KO model takes the form \cite{Kulik1975}
\begin{eqnarray}
    &&\Xi(\varphi_2,T)=\cos{\left( \frac{\varphi_2}{2}\right)}\times\nonumber\\
    &&\int\limits_{\Delta(T)\cos{\left( \frac{\varphi_2}{2}\right)}}^{\Delta(T)}\frac{\tanh{\frac{\varepsilon}{k_BT}}}{\sqrt{\varepsilon^2-\Delta^2(T)\cos^2{\left(\frac{\varphi_2}{2}\right)}}}\mathrm{d}\varepsilon.
\end{eqnarray}
%and $\varepsilon$ is an integration variable. 
In the zero-temperature limit ($T=0$), the KO CPR can be simplified in \cite{Kulik1975}
\begin{equation}
    I_{J_2}(\varphi_2,T=0)=\frac{\pi\Delta_0}{eR_2}\cos{\left(\frac{\varphi_2}{2}\right)}\text{arctanh}\left[ \sin{\left(\frac{\varphi_2}{2}\right)}\right],
    \label{eq:0TIJ}
\end{equation}
where $\Delta_0$ is the zero-temperature superconducting energy gap of the ring.

Figure \ref{Fig1}-b shows the normalized CPR of $J_2$ [$I_{J_2}(T)/I_{C2,0}$, with $I_{C2,0}=\frac{\pi\Delta_0}{eR_2}$ the zero-temperature junction critical current] as a function of $\varphi_2$ for different values of temperature (normalized with respect to the critical temperature $T_C$). By rising $T$, the CPR evolves from a skewed to a perfect sinusoidal phase-dependence \cite{Kulik1975,Golubov2004}. This behavior entails the higher responsivity of $J_2$ when the L-SQUIPT is operated at low temperature. 
Notably, the CPR is almost the same for the $T=0$ limit (black squares) and for $T=0.01T_C$ (cyan line), as shown in Fig. \ref{Fig1}-b.

The resulting temperature-dependent Josephson inductance ($L_{J_2}$) of $J_2$ can be written as
\begin{equation}
    L_{J_2}(\varphi_2, T)=\frac{\hbar}{2e}\frac{\text{d}\varphi_2}{\text{d}I_{J_2}(\varphi_2,T)}.
    \label{KinInd}
\end{equation}
In the zero-temperature limit, the kinetic inductance can be obtained by substituting Eq. \ref{eq:0TIJ} in the above expression. The resulting closed form is therefore
\begin{eqnarray}
        && L_{J_2}(\varphi_2,T=0)=\frac{I_{C2,0}}{2e}/\nonumber \\
        &&\left[-\frac{1}{2}\sin{\left(\frac{\varphi_2}{2}\right)}\operatorname{artanh}\left[\sin{\left(\frac{\varphi_2}{2}\right)}\right]+\frac{\cos^2\left(\frac{\varphi_2}{2}\right)}{2-2\sin^2\left(\frac{\varphi_2}{2}\right)}\right].
        \label{ZTKOL}
\end{eqnarray}
Figure \ref{Fig1}-c shows the normalized Josephson inductance $L_{J2}/L_{J2,0}$ (with $L_{J2,0}$ its zero-temperature and zero-phase value) as a function of $\varphi_2$ calculated for different temperatures. At low temperature ($T\leq0.3T_C$) and for $\varphi_2 \to \pi$, the Josephson inductance drops of about one order of magnitude with respect to $L_{J2,0}$. Furthermore, in the limit of $\varphi_2 \to \pi$, the zero-temperature kinetic inductance (see Eq. \ref{ZTKOL}) goes to zero. As a matter of fact, the vanishing of $L_{J_2}$ would not allow to efficiently phase-bias the Josephson junction in a conventional SQUIPT, since its inductance becomes smaller than that of the ring ($L_2$). As we shall show below, the L-SQUIPT allows to exploit the full phase bias of $J_2$ yielding largely enhanced transfer functions even at the lowest temperatures, where the magnetometer is expected to show its maximum magnetic flux sensitivity.

To perform the read-out operation, the weak-link $J_2$ is equipped with a superconducting readout tunnel-probe ($P$, green electrode in Fig. \ref{Fig1}-a), as in conventional SQUIPTs \cite{Giazotto2011}.
On the one hand, this geometry allows to operate the magnetometer by conventional quasiparticle transport measurements in both voltage and current bias. On the other hand, the $\Phi_1$-dependent Josephson coupling between $J_2$ and $P$ can be employed to design different dissipationless read-out schemes for the L-SQUIPT. In particular, the variation of the Josephson output tunnel junction inductance ($L_{out}$) can be detected by an inductively coupled SQUID read-out or by dispersive microwave measurements.   

\section{Phase-biasing $J_2$}
\label{Phase-Bias}

\begin{figure}[t!]
\begin{center}
    \includegraphics [width=\columnwidth]{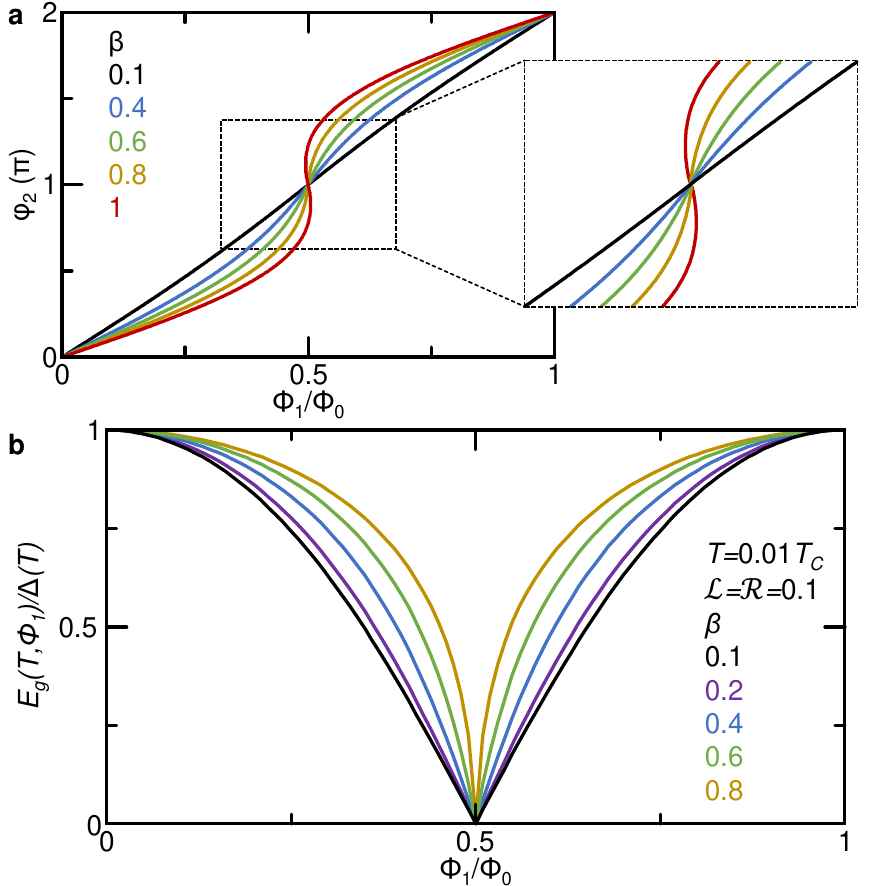}
\end{center}
        \caption{\textbf{(a)} Phase drop ($\varphi_2$) across $J_2$ as a function of the external flux $\Phi_1$ calculated assuming $T=0.01T_C$ and $\mathcal{L}=\mathcal{R}=0.1$ for different values of $\beta$. Inset: blow up of the $\varphi_2$ versus $\Phi_1$ characteristics calculated around $\Phi_1=0.5\Phi_0$. \textbf{(b)} Minigap induced in $J_2$ ($E_g$) as a function of $\Phi_1$ calculated at $\mathcal{L}=\mathcal{R}=0.1$ and $T=0.01T_C$ for different values of $\beta$.}
        \label{Fig2}
\end{figure}

The dependence of the phase drop $\varphi_2$ across $J_2$ on the external magnetic flux $\Phi_1$ can be calculated by considering three conditions typical of Josephson interferometers: (i) the quantization of the magnetic flux piercing the first loop; (ii) the phase-locking between $J_1$ and $J_2$; (iii) the circulating supercurrent conservation in the L-SQUIPT double-loop. 

Thanks to the flux quantization, the phase drop across $J_1$ is related to $\Phi_1$ through 
\begin{equation}
     \varphi_1+\frac{2\pi L_1}{\Phi_0}I_{R}=\frac{2\pi\Phi_1}{\Phi_0},
     \label{Ffconv}
\end{equation}
where $I_R$ is the total supercurrent circulating in the L-SQUIPT, and $\Phi_0=2.067\times 10^{-15}$ Wb is the flux quantum. Equation \ref{Ffconv} describes the $\Phi_1\to\varphi_1$ conversion by taking into account the finite inductance $L_1$ of the first superconducting ring necessary to efficiently couple the L-SQUIPT to the external magnetic field. 
Furthermore, the phase drop $\varphi_2$ across $J_2$ is locked to $\varphi_1$ by the equation
\begin{equation}
          \varphi_2-\varphi_1+\frac{2\pi L_2}{\Phi_0}I_{J_2}=0,
          %\frac{2\pi\Phi_2}{\Phi_0}=0
          \label{eq:phase-locking}
\end{equation}
since the magnetic flux through the second ring is assumed to be zero ($\Phi_2=0$). Equation \ref{eq:phase-locking} illustrates the $\varphi_1 \to \varphi_2$ conversion, which is strongly influenced by the finite inductance of the second ring ($L_2$). 

Finally, to calculate the $\varphi_2(\Phi_1)$ characteristics, we need to consider the conservation of the circulating current in the L-SQUIPT. This implies that $I_{R}$ is distributed between the two Josephson junctions $J_1$ and $J_2$, that is 
\begin{equation}
    I_{R}=I_{J_1}+I_{J_2}.
    \label{eq:current_conservation}
\end{equation}

As a result, the phase drop across $J_2$ as a function of the external flux piercing the first loop reads
\begin{equation}
    \varphi_2=\frac{2\pi \Phi_1}{\Phi_0}- \beta \Big\{\sin{\left[ \varphi_2+\beta\mathcal{L}\mathcal{R}\Xi(\varphi_2)\right]}+\mathcal{R}\Xi(\varphi_2)(1+\mathcal{L})  \Big\},
    \label{eq:phi2}
\end{equation}
where $\beta=(2\pi L_1 I_{C1})/\Phi_0$ is the screening parameter accounting for the finite inductance $L_1$ of the first loop, $\mathcal{L}=L_2/L_1$ describes the difference between the inductance of the two loops, and $\mathcal{R}=I_{C2,0}/I_{C1,0}$ takes into account the asymmetry in the critical current of $J_1$ and $J_2$. We note that to have an efficient $\Phi_1\to\varphi_2$ transduction, the ring inductance of the two loops need to satisfy $L_1\gg L_2$, that is $\beta\ll1$ is required in Eq. \ref{eq:phi2}. 

Figure \ref{Fig2}-a shows the $\varphi_2(\Phi_1)$ characteristics calculated by solving Eq. \ref{eq:phi2} for different values of $\beta$ at $\mathcal{L}=\mathcal{R}=0.1$ and $T=0.01T_C$. When the inductance of the first loop is large, the phase bias shows a strong non-linearity with $\Phi_1$. In particular, $\varphi_2$ shows multiple solutions for $\beta \gtrsim0.8$ in the flux range $\Phi_1\to \Phi_0$ thereby preventing to fully phase-bias $J_2$. By contrast, for lower values of $\beta$, the phase drop is a continuous function of the external magnetic flux, thus $J_2$ is sensitive to each value of $\Phi_1$. In the following, we will use $\beta=0.8$ to optimize the phase-bias of $J_2$ and, therefore, to maximize the sensitivity of the L-SQUIPT magnetometer. We note that $J_2$ shows a vanishing inductance in these conditions, since it operates in the \textit{short}-junction limit and at low temperatures (see Fig. \ref{Fig1}-b). Therefore, an efficient phase-bias of the junction would be impossible in a conventional SQUIPT. 

The $\Phi_1$-dependent values of $\varphi_2$ strongly influence the DoS ($\mathcal{N}_{J2}$) of the normal metal element forming the output Josephson junction. Since $J_2$ is assumed to be in the \textit{short}-junction limit, its DoS takes the form \cite{Artemenko1979,Heikkila2002}
\begin{equation}
        \begin{split}
        \mathcal{N}_{J2}(x,\epsilon,\varphi_2)=\text{Re}\sqrt{\frac{(\epsilon+i\Gamma)^2}{(\epsilon+i\Gamma)^2-\Delta^2(T)\cos^2(\varphi_2/2)}}\times\\
        \cosh\left( \frac{2x-L}{L}\text{arccosh}\sqrt{\frac{(\epsilon+i\Gamma)^2-\Delta^2(T)\cos^2(\varphi_2/2)}{(\epsilon+i\Gamma)^2-\Delta^2(T)}}\right),
        \end{split}
        \label{eq:DoS}
\end{equation}
where $\epsilon$ is the energy relative to the chemical potential of the superconductors, $\Gamma$ is the Dynes broadening parameter \cite{Dynes1984} and $x\in [0,L]$ is the spatial coordinate along the $J_2$ length. Equation \ref{eq:DoS} highlights that the density of states is strongly tuned by $\varphi_2$. In particular, the superconducting minigap induced in N by the proximity to S \cite{DeGennes} takes the form 
\begin{equation}
    E_g(T,\varphi_2)=\Delta(T)\cos{\left(\frac{\varphi_2}{2}\right)}.
    \label{eq:minigap}
\end{equation}
We note that the value of $E_g$ is constant along the nanowire length. For $\varphi_2=0$ the induced minigap is maximum [$E_g(T,0)=\Delta(T)$], while for $\varphi_2=\pi$ the nanowire shows the normal metal DoS [$E_g(T,\pi)=0$]. 

In the L-SQUIPT, the dependence of the minigap on the external magnetic flux [$E_g(T,\Phi_1)$] can be calculated by combining Eqs. \ref{eq:phi2} and \ref{eq:minigap}. Figure \ref{Fig2}-b presents the dependence of $E_g(T)$ on $\Phi_1$ calculated at $T=0.01T_C$ and $\mathcal{L}=\mathcal{R}=0.1$ for different values of $\beta$. By enhancing the screening parameter, the minigap shows a stronger variation with external magnetic flux at $\Phi_1\to0.5\Phi_0$. On the contrary, the minigap is more sensitive at $\Phi_1\to\Phi_0$ for low values of $\beta$, but its maximum steepness is limited. Therefore, the L-SQUIPT magnetometer is expected to show higher sensitivity for large values of $\beta$ at $\Phi_1\to0.5\Phi_0$. 

\begin{figure}[t!]
\begin{center}
    \includegraphics [width=\columnwidth]{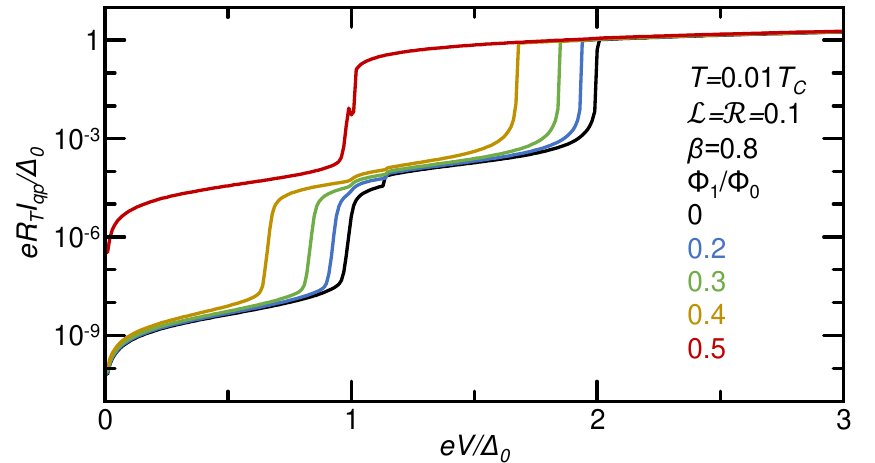}
\end{center}
        \caption{Quasiparticle current as a function of voltage calculated at $\beta=0.8$ and $\mathcal{L}=\mathcal{R}=0.1$, $T=0.01T_C$ for several values of $\Phi_1$. All the curves assume $\Gamma=10^{-4}\Delta_0$.}
        \label{Fig3}
\end{figure}

\section{Dissipative read-out}
\label{Dissipative}
Here, we discuss the magnetic flux dependent quasiparticle transport between $J_2$ and $P$. This will allow us to evaluate the sensitivity of the L-SQUIPT magnetometer both in the voltage-bias and current-bias configurations.

\subsection{Quasiparticle transport}

The quasiparticle current flowing through the $J_2$-$P$ tunnel junction can be written as \cite{Giazotto2011}

\begin{equation}
        I_{qp}=\frac{1}{ewR_T }\int\limits_{\frac{L-w}{2}}^{\frac{L+w}{2}}\text{d}x\int\limits_{-\infty}^{\infty}\text{d}\epsilon\mathcal{N}_{J2}(x,\epsilon,\varphi_2)\mathcal{N}_{p}(\epsilon, V)\mathcal{F}(\epsilon,V),
\end{equation}
where $R_T$ and $w$ are the normal-state resistance and the width of the junction, respectively. 
Furthermore, $\mathcal{F}(\epsilon,V)=[f_0(\epsilon-eV)-f_0(\epsilon)]$ is the difference between the Fermi-Dirac distribution functions ($f_0$) of the two electrodes.
The normalized Bardeen-Cooper-Schrieffer (BCS) DoS of the superconducting tunnel probe can be written as
\begin{equation}
\mathcal{N}_{p}(\epsilon,V)=\left|\text{Re}\left[\frac{(\epsilon-eV+i\Gamma)}{\sqrt{(\epsilon-eV+i\Gamma)^2-\Delta^2(T)}}\right]\right|.
\end{equation}
For simplicity, we assume the $P$ to be made of the same superconductor of the L-SQUIPT ring.

The typical quasiparticle current ($I_{qp}$) versus voltage ($V$) characteristics of the L-SQUIPT calculated assuming $\beta=0.8$, $\mathcal{L}=\mathcal{R}=0.1$ and $T=0.01T_C$ are shown in Fig. \ref{Fig3} for several values of the external magnetic flux. The quasiparticle current tunnels through the barrier when the voltage bias is larger than the sum of the energy gaps of $J_2$ and $P$, that is for $eV\geq\Delta(T)+E_g(T,\Phi_1)$ \cite{tinkham2012introduction}. Indeed, the threshold voltage is maximal for $\Phi_1=0$ (black curve), since the minigap in $N$ acquires the same value of the energy gap of the superconducting ring [$E_g(T,0)=\Delta(T)$]. By rising the external magnetic flux, large quasiparticle tunneling occurs at lower values of voltage bias until reaching its minimum value $eV=\Delta(T)$ for $\Phi_1=0.5\Phi_0$ [red curve, since $E_g(T,0.5\Phi_0)=0$]. The variation of the $I_{qp}(V)$ characteristics with the magnetic flux is stronger in the interval $0.4\Phi_0\leq\Phi_1\leq0.5\Phi_0$, since $E_g$ shows a stark dependence on $\Phi_1$ in this range (see Fig. \ref{Fig2}-b). As a consequence, the L-SQUIPT can be operated as a sensitive magnetometer by simple measurements of the output junction voltage in current bias mode or the tunneling quasiparticle current in voltage bias.  

\begin{figure}[t!]
\begin{center}
    \includegraphics [width=\columnwidth]{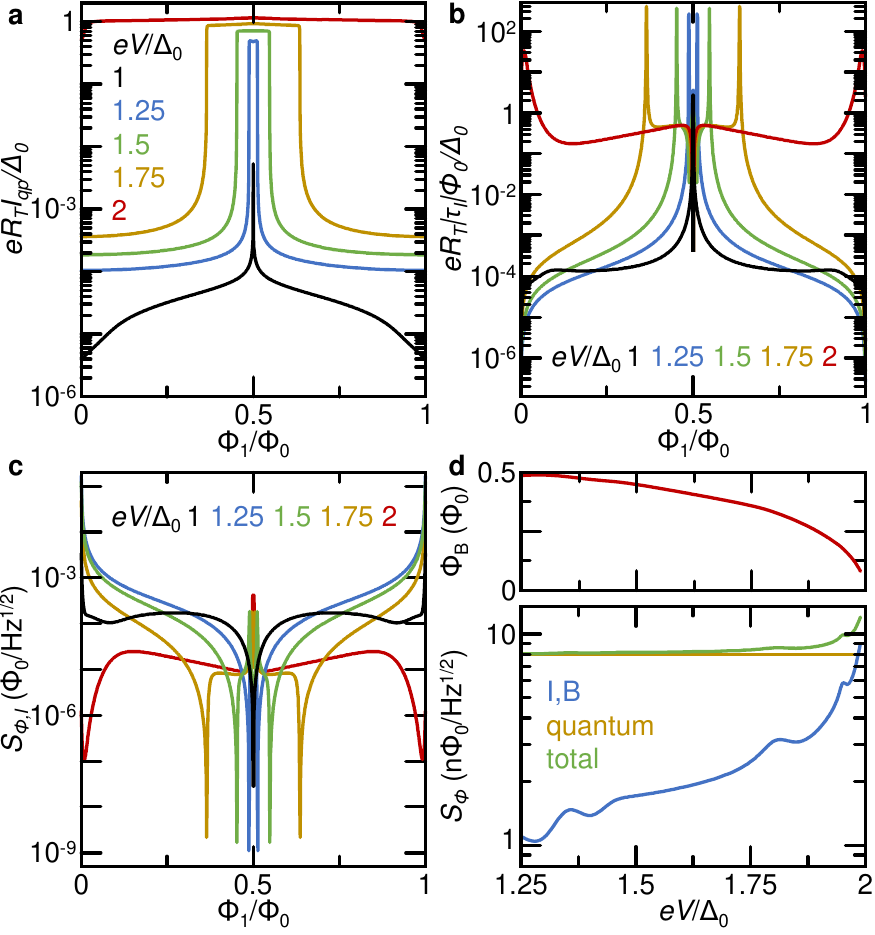}
\end{center}
        \caption{\textbf{Voltage bias operation.} \textbf{(a)} Quasiparticle current as a function of the external magnetic flux calculated for several values of $V$. \textbf{(b)} Flux-to-current transfer function versus $\Phi_1$ calculated for different values of bias voltage. \textbf{(c)} Flux sensitivity per unit bandwidth as a function of $\Phi_1$ for different values of $V$. In the calculations we set $R_T=50$ k$\Omega$ and $\Delta_0=200\;\mu$eV. 
        \textbf{(d)} Top: Magnetic flux corresponding to best sensitivity versus $V$. Bottom: best flux sensitivity per unit bandwidth (blue) and quantum noise (gold) versus $V$. The total sensitivity is shown in green. 
        All the panels assume $\beta=0.8$, $\mathcal{L}=\mathcal{R}=0.1$, $T=0.01T_C$ and $\Gamma=10^{-4}\Delta_0$.}
        \label{Fig4}
\end{figure}

\subsection{Voltage bias operation}
The voltage bias operation of the L-SQUIPT magnetometer takes advantage of the strong $\Phi_1$ dependence of $I_{qp}$ for specific values of $V$, as shown in Fig. \ref{Fig4}-a. In particular, the current is almost independent of the magnetic flux for $V=2\Delta_0/e$, since the output junction is always biased in the normal-state. For lower values of the bias voltage, the modulation of $E_g$ with $\Phi_1$ results in the strong variation of $I_{qp}$ with the magnetic flux. Indeed, by decreasing $V$ the maximum steepness of the curves moves towards $\Phi_1=0.5\Phi_0$, that is reached for $V=\Delta_0/e$ thus  corresponding to $E_g(\Phi_1)=0$ (see Fig. \ref{Fig2}-b). As a consequence, the voltage bias operation of the L-SQUIPT requires $\Delta_0<eV<2\Delta_0$.

The $\Phi_1$ dependence of $I_{qp}$ is completely reflected in the flux-to-current transfer function, which is defined as
\begin{equation}
    \tau_I=\frac{\text{d}I_{qp}}{\text{d}\Phi_1}.
\end{equation}
Figure \ref{Fig4}-b shows $\tau_I$ versus $\Phi_1$ calculated for the same parameters of $I_{qp}$ (panel a). At a given $V$, the maximum value of the transfer function corresponds to the strongest variation of $I_{qp}$ with $\Phi_1$, while $\tau_I=0$ for $\Phi_1=0.5\Phi_0$ (where $I_{qp}$ shows its maximum, see Fig. \ref{Fig4}-a).

The most common figure of merit for a magnetometer is the the flux noise, i.e., the flux sensitivity per unit bandwidth. In voltage bias operation, this can be written as
\begin{equation}
        S_{\Phi,I}=\frac{\sqrt{S_I}}{|\tau_I|},
\end{equation}
where $S_I$ is the current-noise spectral density. The latter reads
\begin{equation}
        S_I=2eI_{qp}(V)\coth{\frac{eV}{2k_BT}}.
\end{equation}
Figure \ref{Fig4}-c shows the $\Phi_1$ dependence of $S_{\Phi,I}$ for the L-SQUIPT calculated at different values of bias voltage. For these simulations we assume a geometry and materials feasible by standard fabrication techniques. Indeed, we set $\Delta_0=200\;\mu$eV (aluminum) for the superconducting ring and the output tunnel probe while considering a tunnel resistance $R_T=50$ k$\Omega$. The flux sensitivity strongly depends on both $\Phi_1$ and $V$. Indeed, depending on the magnetic flux of interest, the best operating point ($\Phi_B$) can be chosen by tuning the bias voltage (see the top panel of Fig. \ref{Fig4}-d). The flux noise corresponding to $\Phi_B$ is $S_{\Phi, I_{best}}<10$ n$\Phi_0/\sqrt{\text{Hz}}$ for a bias voltage in the range $1.25\Delta_0\leq eV<2\Delta_0$. We note that, in a superconducting interferometer, the ultimate flux sensitivity is limited by the quantum noise ($S_{\Phi,q}$) defined as \cite{Kirtley2010}
\begin{equation}
        S_{\Phi,q}=\sqrt{\hbar L_1}.
\end{equation}
By substituting $I_{C1}=100\;\mu$A in the screening parameter equation, we obtain $L_1=2.6$ pH. The resulting quantum noise due to the inductance of the superconducting ring is $S_{\Phi,q}=8$ n$\Phi_0/\sqrt{\text{Hz}}$. Therefore, the L-SQUIPT magnetometer operated in voltage bias shows a quantum limited flux sensitivity for most values of $V$. In fact, $S_{\Phi,q}$ dominates the total flux noise of the device, defined as $S_{\Phi,t}=\sqrt{s_{\Phi,I}^2+S_{\Phi,q}^2}$, in almost the full voltage range (see bottom panel of Fig. \ref{Fig4}-d).

\subsection{Current bias operation}

\begin{figure}[t!]
\begin{center}
    \includegraphics [width=\columnwidth]{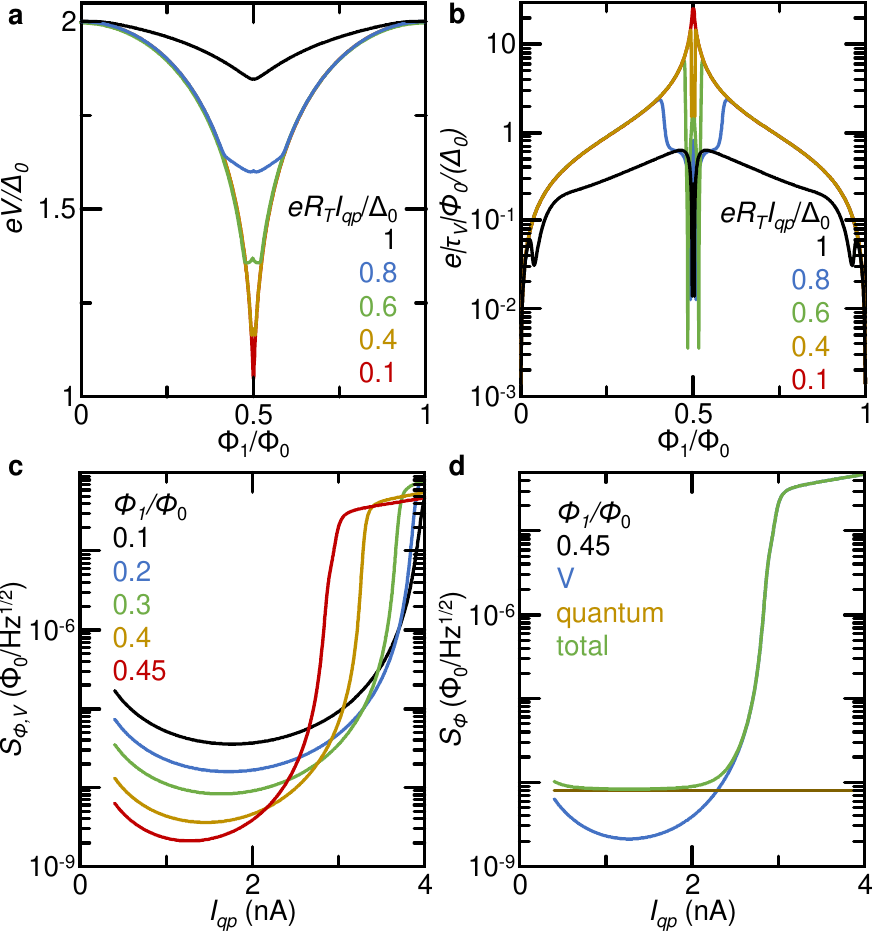}
\end{center}
        \caption{\textbf{Current bias operation.} \textbf{(a)} Output voltage as a function of the external magnetic flux calculated for several values of $I_{qp}$. \textbf{(b)} Flux-to-voltage transfer function versus $\Phi_1$ calculated for different values of bias current. \textbf{(c)} Flux sensitivity per unit bandwidth as a function of $I_{qp}$ for different values of $\Phi_1$. In the calculations we employed $R_T=50$ k$\Omega$ and $\Delta_0=200\;\mu$eV. 
        \textbf{(d)} Flux sensitivity per unit bandwidth (blue) and quantum noise (gold) versus $I_{qp}$ calculated for $\Phi_1=0.45\Phi_0$. The total sensitivity is shown in green. 
        All the panels assume $\beta=0.8$, $\mathcal{L}=\mathcal{R}=0.1$, $T=0.01T_C$ and $\Gamma=10^{-4}\Delta_0$.}
        \label{Fig5}
\end{figure}

The current bias operation of the L-SQUIPT magnetometer exploits the dependence of $V$ on $\Phi_1$ while a constant $I_{qp}$ is injected in the device, as shown in Fig. \ref{Fig5}-a. For $I_{qp}=\Delta_0/(eR_T)$, the modulation of $V$ with the external magnetic flux is limited, while by decreasing the bias current the voltage span and the steepness of the curves increase. This behavior is highlighted by the flux-to-voltage transfer function
\begin{equation}
    \tau_V=\frac{\text{d}V}{\text{d}\Phi_1}.
\end{equation}
Indeed, the maximum value of the transfer function rises while moving towards $0.5\Phi_0$ by increasing the bias current (see Fig. \ref{Fig5}-b). In particular, the L-SQUIPT shows $\tau_V\simeq26\Delta_0/(e\Phi_0)$ at $\Phi_1=0.498\Phi_0$ for $I_{qp}=0.1\Delta_0/(eR_T)$. 

In current bias operation, the flux sensitivity per unit bandwidth can be written as
\begin{equation}
        S_{\Phi,V}=\frac{\sqrt{S_V}}{|\tau_V|},
\end{equation}
where the voltage-noise spectral density takes the form
\begin{equation}
        S_{V}=R_d^2S_I=\left(\frac{\text{d}V}{\text{d}I}\right)^2S_I.
\end{equation}
In the above equation, $R_d=\text{d}V/\text{d}I$ is the differential resistance of the output Josephson junction. Figure \ref{Fig5}-c shows $ S_{\Phi,V}$ as a function of $I_{qp}$ for several values of $\Phi_1$. 
For the simulations we considered the same structure of the voltage bias configuration, that is $\Delta_0=200\;\mu$eV for both the superconducting ring and the output tunnel electrode ($R_T=50$ k$\Omega$). The L-SQUIPT sensitivity is maximum for high value of magnetic flux, as a result of the increased flux-to-voltage transfer function (see Fig. \ref{Fig5}-b). In particular, a flux sensitivity of $\sim2$ n$\Phi_0/\sqrt{\text{Hz}}$ can be reached for $\Phi=0.45\Phi_0$ at $I_{qp}\simeq1.3$ nA. 

Also in the current bias mode, the L-SQUIPT magnetometer shows a quantum limited flux sensitivity. Indeed, $S_{\Phi,q}=8$ n$\Phi_0/\sqrt{\text{Hz}}$ dominates the total flux noise of the device in a wide range of bias currents, as shown in Fig. \ref{Fig5}-d for $\Phi=0.45\Phi_0$.

\subsection{Temperature dependence}
\begin{figure}[t!]
\begin{center}
    \includegraphics [width=\columnwidth]{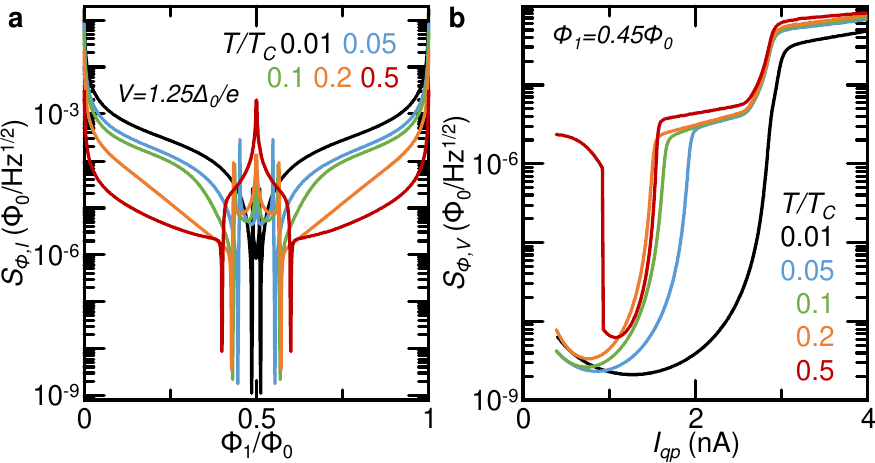}
\end{center}
        \caption{
        \textbf{Temperature dependence of the flux sensitivity.} \textbf{(a)} Flux sensitivity per unit bandwidth as a function of $\Phi_1$ for different values of temperature calculated at $V=1.25\Delta_0/e$.  \textbf{(b)} Flux sensitivity per unit bandwidth as a function of $I_{qp}$ calculated for different values of temperature at $\Phi_1=0.45\Phi_0$.
        All the panels assume $\beta=0.8$, $\mathcal{L}=\mathcal{R}=0.1$, $\Gamma=10^{-4}\Delta_0$, $R_T=50$ k$\Omega$ and $\Delta_0=200\;\mu$eV.}
        \label{Fig6}
\end{figure}

Here, we investigate the temperature dependence of the L-SQUIPT performance both in voltage and in current bias. To this end, we set the same device parameters of previous sections, that is $\mathcal{L}=\mathcal{R}=0.1$, $\Gamma=10^{-4}\Delta_0$, $R_T=50$ k$\Omega$ and $\Delta_0=200\;\mu$eV. Since $\beta$ is a temperature dependent parameter, we consider $\beta(T=0.01T_C)=0.8$ taking into account the exponential damping of $I_{C1}$ with $T$ \cite{Golubov2004}. Indeed, Eq \ref{AmbeEq} can be employed for $T\geq0.05T_C$ in an aluminum/copper SNS junction of length $L=2\;\mu$m (with $D=6\times10^{-3}$m$^2$s$^{-1}$ the diffusion coefficient of copper).

Figure \ref{Fig6}-a shows the flux sensitivity in the voltage bias operation (at $V=1.25\Delta_0/e$) as a function of $\Phi_1$ calculated for different values of $T$. By increasing the temperature, the best value of $S_{\Phi,I}$ rises substantially, while $\Phi_B$ is only slightly affected from $T$. On the contrary, the sensitivity far way from the best operating point improves by increasing temperature, since the flux-to-current transfer function shows a smoother dependence in $\Phi_1$. Indeed, at high temperature, the CPR of $J_2$ shows a lower slope around $\varphi_2=\pi$ (see Fig. \ref{Fig1}-b), thus causing a smaller variation of $I_{qp}$ with $\Phi_1$. This degrades the best performance of the L-SQUIPT, but it provides a more constant sensitivity in the whole magnetic flux range (see the red curve in Fig. \ref{Fig6}-a).

Figure \ref{Fig6}-b shows the temperature dependence of the flux sensitivity of the L-SQUIPT operated in current bias at $\Phi_1=0.45\Phi_0$. By rising the temperature, the magnetometer best sensitivity is slightly affected by temperature, but the range fo bias current showing high sensitivity narrows. For $T=0.5T_C$, a flux sensitivity $S_{\Phi,V}\sim10$ n$\Phi_0/\sqrt{\text{Hz}}$ is obtained for $I_{qp}\sim1$ nA. Thus, the L-SQUIPT is a quantum limited magnetometer both in voltage and current bias operations only for $T<0.5T_C$.

\section{Dissipationless read-out}
\label{LessRead}
Here, we discuss the magnetic flux dependent Josephson transport between $J_2$ and $P$. This will allow us to evaluate the sensitivity of the L-SQUIPT magnetometer in different dissipationless read-out geometries.

\subsection{Josephson current and inductance}
Since we assume a superconducting tunnel read-out probe, a dissipationless zero-bias current ($I_{out}$) can flow thanks to Josephson coupling. The latter can be calculated by means of the Ambegaokar-Baratoff equation for a point-like junction \cite{Giazotto2011,Ambegaokar1963}
\begin{equation}
    \begin{split}
        I_{out}(T,\varphi_2)=\frac{\pi E_g(T,\varphi_2) \Delta (T) k_B T}{eR_T}\times\\
        \sum_{l=0,\pm1,...}\frac{1}{\sqrt{[\omega_l^2+E_g^2(T,\varphi_2)][\omega_l^2+\Delta^2(T)]}},
    \end{split}
    \label{Outcurr}
\end{equation}
where $\omega_l=\pi k_b T(2l+1)$. 

\begin{figure}[t!]
\begin{center}
    \includegraphics [width=\columnwidth]{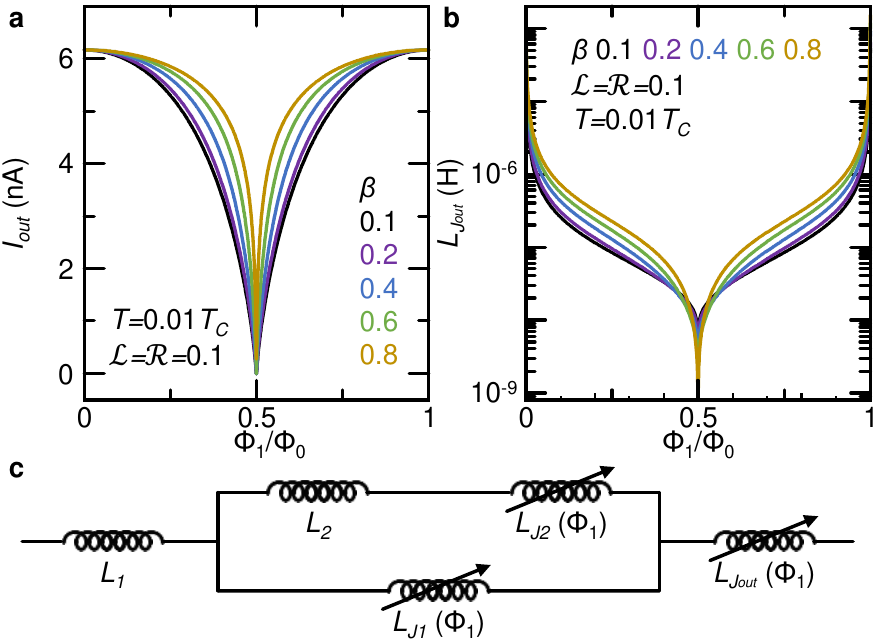}
\end{center}
        \caption{
        \textbf{Josephson transport through the read-out junction.} \textbf{(a)} Josephson current flowing through the output tunnel junction as a function of $\Phi_1$ calculated for different values of $\beta$. 
        \textbf{(b)}  Kinetic inductance of the output tunnel junction as a function of $\Phi_1$ calculated for different values of $\beta$. Both panels assume $\mathcal{L}=\mathcal{R}=0.1$, $\Gamma=10^{-4}\Delta_0$, $T=0.01T_C$, $R_T=50$ k$\Omega$ and $\Delta_0=200\;\mu$eV.
        \textbf{(c)} Schematic representing all the components of the total inductance ($L_{TOT}$) of the L-SQUIPT. We represented all the inductance contributions that directly or indirectly depend on $\Phi_1$. 
        }
        \label{Fig7}
\end{figure}

Figure \ref{Fig7}-a shows $I_{out}$ versus $\Phi_1$ calculated for a L-SQUIPT at $T=0.01~T_C$ and $\mathcal{L}=\mathcal{R}=0.1$ for different values of $\beta$. 
We consider the same structure of the previous calculations, that is $\Delta_0=200\;\mu$eV and $R_T=50$ k$\Omega$. At $\Phi_1=0$, all the curves collapse to the same maximum value, that is the output junction critical current ($I_{C,out}\simeq6.2$ nA). By rising the external magnetic flux, $I_{out}$ lowers until reaching its minimum at $\Phi_1=0.5\Phi_0$, because the superconducting energy gap of $J_2$ closes. The energy gap shows a steeper dependence on $\Phi_1$ for large values of $\beta$ (see Fig. \ref{Fig2}-b). This behavior is transferred to the Josephson current. Indeed, the variation of $I_{out}$ around $\Phi_1=0.5~\Phi_0$ is less sharp for low values of the screening parameter (see Fig. \ref{Fig7}-a).

The suppression of the critical current causes the variation of the Josephson inductance of the read-out junction ($L_{J_{out}}$). The latter can be calculated through Eq. \ref{KinInd} by considering the phase of the tunnel probe equal to 0. Consequently, $L_{J_{out}}$ is proportional to the derivative of $\varphi_2$ with respect to $I_{out}$. Figure \ref{Fig7}-b shows $L_{J_{out}}$ versus $\Phi_1$ calculated starting from the Josephson currents shown in panel a. For all values of $\beta$, the inductance spans over several orders of magnitude. By increasing $\beta$, the overall variation of $L_{J_{out}}$ with the magnetic flux increase and its maximum steepness moves from $\Phi_1\to 0$ to $\Phi_1\to 0.5\Phi_0$. 

For the implementation of dissipationless read-out schemes, we need to consider the total inductance of the L-SQUIPT ($L_{TOT}$) represented in Fig. \ref{Fig7}c. 
The small ring inductance $L_2$ and the second junction inductance $L_{J_2}$ are in series. This block is in parallel with the inductance ($L_{J_1}$) of the junction $J_1$. The resulting inductance is in series with the large loop inductance $L_1$.
All these are in series with the Josephson output tunnel junction inductance $L_{out}$.
Thus, the total inductance is 
\begin{equation}
    L_{TOT}(\Phi_1)= L_{1} + \frac{L_{J_1}(\Phi_1) [L_2 +L_{J_2}(\Phi_1)]}{L_2+ L_{J_1}(\Phi_1) +L_{J_2}(\Phi_1)} + L_{J_{out}}(\Phi_1).
\end{equation}

The variation of Josephson inductance, and thus $L_{TOT}$, can be revealed through different dissipationless read-out schemes. Indeed, the L-SQUIPT can be inductively coupled to a SQUID amplifier, as routinely realized for kinetic inductance detectors (KIDs). Alternatively, the device can be integrated in a $RLC$ resonant circuit, whose resonance frequency varies with $\Phi_1$. 

\subsection{Inductive read-out}

The scheme for the inductive read-out of the L-SQUIPT is shown in Fig. \ref{Fig8}-a. The parallel connection of the L-SQUIPT (of inductance $L_{TOT}$) and a load inductor ($L$) is biased by means of a dc current current generator ($I_b$). Indeed, the variation of $L_{TOT}$ generates a change of the current flowing through the L-SQUIPT ($I_{out}$) and the load resistor ($I_L$). The latter is detected by a dc SQUID inductively coupled to the circuit through the mutual inductance $M$, where the magnetic flux piercing the SQUID is $\Phi_s=MI_L$. For small variations of $\Phi_S$ (linear response regime, $LI_L\ll \Phi_0$) \cite{Giazotto2008}, the current flowing through the load inductor reads 
\begin{equation}
   I_L=I_b\frac{\Phi_0}{\Phi_0+2\pi LI_{out}}.
   \label{loadcurr}
\end{equation}
In this configuration, the L-SQUIPT operates as a magnetic-flux-to-magnetic-flux transducer or magnetic flux amplifier, where the efficiency can be quantify by
\begin{equation}
   \frac{\text{d}\Phi_S}{\text{d}\Phi_1}=\frac{\text{d}I_{out}}{\text{d}\Phi_1}M\frac{\text{d}I_L}{\text{d}I_{out}}.
\end{equation}
The ratio $\text{d}I_{out}/\text{d}\Phi_1$ can be calculated through the derivative of Eq. \ref{Outcurr} with respect of the input magnetic flux. Accordingly, the term $\text{d}I_{L}/\text{d}I_{out}$ can be calculated by performing the $\text{d}I_{out}$ derivative of Eq. \ref{loadcurr}, thus obtaining the following expression
\begin{equation}
   \frac{\text{d}I_L}{\text{d}I_{out}}=2\pi LI_b\frac{\Phi_0}{\left(\Phi_0+2 \pi LI_{out}\right)^2}.
\end{equation}

%%%%%%%%%%%%%%%%%%%%%%%%%%%%%%%%%%%%%%%%%%%%%%%%%%%%%%%%%%%%%%%%%%%%%

\subsection{Dispersive measurement}
\label{sec:dispersive_measurement}
\begin{figure}[t!]
\begin{center}
    \includegraphics [width=\columnwidth]{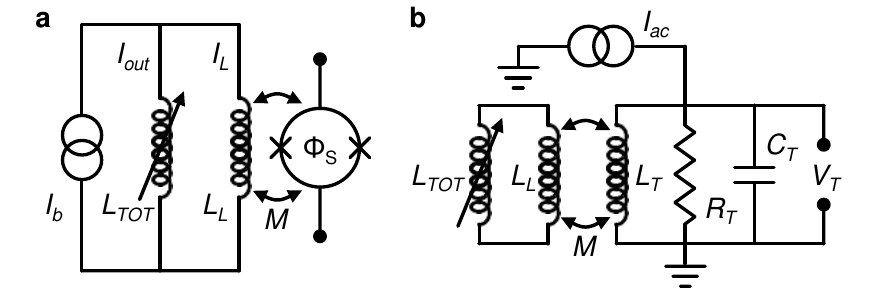}
\end{center}
        \caption{
        \textbf{Dissipationless read-outs for the L-SQUIPT.} \textbf{(a)} Inductive read-out scheme, where the changes of the L-SQUIPT inductance ($L_{TOT}$) are recorded by dc SQUID magnetometer coupled to the device by a mutual inductance $M$. 
        \textbf{(b)} Dispersive read-out scheme, where the changes of $L_{TOT}$ are recorded thanks to the changes of the characteristic frequency of a resonant circuit inductively coupled to the L-SQUIPT.}
        \label{Fig8}
\end{figure}

The scheme for the dispersive read-out of the L-SQUIPT is shown in Fig. \ref{Fig8}-b, where the $\Phi_1$-dependent variation of the Josephson inductance $L_{J_{out}}$ is determined by measuring the resonance frequency of a suited $RLC$ circuit. To this end, a load inductance $L_L$ is added in parallel to the total inductance of the L-SQUIPT ($L_{TOT}$). This circuit is coupled by a mutual inductance ($M$) with a tank circuit characterized by inductance $L_T$, capacitance $C_T$ and resistance $R_T$.
The resulting effective inductance of the tank circuit is \cite{BaronePaterno1982, guarcello2018} 
\begin{equation}
    \tilde{L}_T = L_T \Big( 1- \frac{M^2}{L_T} \frac{1}{L_{TOT}(\Phi_1)+L_L}\Big).
\end{equation}
As a consequence, the resonance frequency of the tank circuit $\tilde{f}_T=1/2\pi \sqrt{\tilde{L}_TC_T}$ strongly depends on the flux $\Phi_1$. 

\subsection{L-SQUIPT noise in dissipationless read-out operation}
We assume that the read-out circuit in Fig. \ref{Fig8} has negligible noise so that the overall sensitivity of L-SQUIPTis determined by its intrinsic noise associated to each element of the device and their correlations. 

From Fig. \ref{Fig8} we can see that noise is essentially determined by the fluctuations of the current across the load inductance $L_L$ that is then coupled my mutual inductance with the read-out circuit.
The current of interest is the tunnel junction one $I_{out}$ in Eq. \ref{Outcurr}.
This is composed by a Johnson thermal contribution (related to the resistances of Josephson junctions) and a phase contribution.

The first one is dominated by the tunnel resistance $R_T$ that is much larger than $R_{J_1}$ and $R_{J_2}$.
For a frequency bandwidth $\delta \omega/(2 \pi)$, it generates a noise voltage of means square value \cite{BaronePaterno1982}
\begin{equation}
    \average{\delta V_{J_{out}}^2} = 4 k_B T R_T \frac{\delta \omega}{ 2 \pi}.
\end{equation}
%Assuming that in a SNS Josephson junction the capacitance is negligible, $J_{out}$ can be treated as a parallel of the resistance $R_T$ and a Josephson inductance $L_{J_{out}}$ \cite{BaronePaterno1982, tinkham2012introduction}.
The tunnel junction can be e represented by an RLC parallel circuit with resistance $R_T$, a Josephson inductance $L_{J_{out}}$ and a small capacitance $C_{out}$.
Passing to the frequency dependent impedances, i.e., $Z_{R_T}=R_T$, $Z_{L_{J_{out}}}= i \omega L_{J_{out}}$, and $Z_{C_{J_{out}}}=1/(i \omega C_{out})$, for the junction $J_{out}$ we have an impedance
\begin{equation}
    \frac{1}{Z_{J_{out}}} =
    \frac{1}{Z_{R_T}}+
    \frac{1}{Z_{L_{J_{out}}}}+
    \frac{1}{Z_{C_{J_{out}}}}.
    \label{eq:z_junction}
\end{equation}
%\begin{equation}
%    Z_{J_{out}} = \frac{Z_{L_{J_{out}}} Z_{R_T}}{Z_{L_{J_{out}}}+ Z_{R_T}}.
%    \label{eq:z_junction}
%\end{equation}

As a consequence, the mean square of the current noise due to the resistance can be written \cite{BaronePaterno1982, solinas2012}
\begin{equation}
    \average{\delta I_{out,R}^2}= \frac{\average{\delta V_{J_{out}}^2}}{|Z_{J_{out}}|^2} = \frac{4 k_B T R_T}{|Z_{J_{out}}|^2} \frac{\delta \omega}{ 2 \pi}.
    \label{eq:deltaI_out}
\end{equation}

The phase noise contribution to $I_{out}$ is due to the fact that, as seen in  Eq. \ref{Outcurr}, $I_{out}$ depends on the phase $\varphi_2$. To calculate it, we first consider the Johnson noise generated by the two separate junctions $J_1$ and $J_2$ at temperature $T$.
Analogously, to what done for the tunnel junction, we have that the noise is
\begin{equation}
    \average{\delta I_{J_i}^2}= \frac{\average{\delta V_{J_i}^2}}{|Z_{J_i}|^2} = \frac{4 k_B T R_{J_i}}{|Z_{J_i}|^2} \frac{\delta \omega}{ 2 \pi}.
    \label{eq:deltaI_1}
\end{equation}
with $i=1,2$ and  $Z_{J_i}$ is obtained by Eq. \ref{eq:z_junction} with the index exchanges.

These uncorrelated noise sources are conbined in the following way.
The small loop can be treated as a SQUID with two Josephson junctions in parallel.
Following Ref. \cite{BaronePaterno1982}, we consider the circulating current $I_{circ}$ as a function of the current flowing through the junction $J_1$ and $J_2$: $I_{circ} = I_{J_1}  - I_{J_2} $.

The fluctuations of the circular current coupled with the loop impedance $L_2$ and determines the phase fluctuations $\delta \varphi_2$. They can be written as $\average{\delta I_{circ}^2}=  \average{\delta I_{J_1}^2} + \average{\delta I_{J_2}^2}$.
By multiplying this expression by $4 \pi^2 L_{2}^2/\Phi_0^2$ and using Eq. \ref{eq:deltaI_1}, we obtain the mean square of the phase noise
\begin{equation}
    \average{\delta \varphi_2^2}= \frac{16 \pi^2 k_B T L_{2}^2 }{\Phi_0^2 } 
    \Big( \frac{R_{J_1}}{|Z_{J_1}|^2} + \frac{R_{J_2}}{|Z_{J_2}|^2} \Big) \frac{\delta \omega}{ 2 \pi}.
    \label{eq:delta_varphi_2}
\end{equation}

The noise current of the tunnel junction is sum of the square of the thermal and phase noise (Eqs. \ref{eq:deltaI_out} and \ref{eq:delta_varphi_2}, respectively)
\begin{equation}
    \average{\delta I_{out}^2} =   \frac{4 k_B T R_T}{|Z_{J_{out}}|^2} \frac{\delta \omega}{ 2 \pi} + c_{\varphi_2} \average{\delta \varphi_2^2}.
    \label{noisecurr}
\end{equation}
The coefficient $c_{\varphi_2}$ is obtained from Eqs. \ref{eq:minigap} and \ref{Outcurr} taking a small variation of $I_{out}$ a a function of $\varphi_2$
\begin{eqnarray}
    c_{\varphi_2} &=&         \frac{1}{4} \left(\frac{\pi \Delta^2 (T) k_B T}{eR_T} \right)^2 \sin^2 \frac{\varphi_2}{2} \nonumber \\
        && \Big[\sum_{l}\frac{1}{\sqrt{[\omega_l^2+E_g^2(T,\varphi_2)][\omega_l^2+\Delta^2(T)]}} \Big]^2.
        \label{coeffpahse}
\end{eqnarray}

Finally, the current fluctuations across $L_L$ induce flux fluctuations 
\begin{equation}
    \average{\delta \Phi_L^2} = \average{\delta I_{out}^2} L_L^2.
\end{equation}
This is the intrinsic magnetic flux noise of the L-SQUIPT that is the ultimate limit the sensitivity of the measurement through the read-out circuit in Fig. \ref{Fig8}. 

We note that the performance for the dissipationless read-out of a L-SQUIPT depend non-trivially on $R_T$, $\delta \omega$ and $\Phi_1$. On the one hand, increase of $R_T$ and $\delta \omega$ increase both contributions of $\average{\delta I_{out}^2}$ (see Eq. \ref{noisecurr}). On the other hand, large values of $R_T$ and $\delta \omega$ increase $Z_{J_{out}}$ (see Eq. \ref{eq:z_junction}) thus suppressing the current noise. 
Similarly, $\Phi_1$ acts on several quantities in Eq. \ref{noisecurr} with unexpected consequences. In particular, the noise is minimum for $\Phi_1 \to 0$ ($\varphi_2 \to 0$), since phase noise contribution becomes negligibly small, as shown by Eq. \ref{coeffpahse}. The best magnetic flux fluctuation for a L-SQUIPT operated in dissipationless mode is $\average{\delta \Phi_L}\sim 4\;\mu\Phi_0$ obtained at $\Phi_1=0$ for
$\mathcal{L}=\mathcal{R}=0.1$, $\Gamma=10^{-4}\Delta_0$, $T=20$ mK, $R_T=50$ k$\Omega$, $\Delta_0=200\;\mu$eV, $\delta \omega/2\pi=100$ MHz, $C_{J_{out}}=1$ fF and $L_L=10$ pH. 

%%%%%%%%%%%%%%%%%%%%%%%%%%%%%%%%%%%%%%%%%%%%%%%%%%%%%%%%%%%%%%%%%%%%%
\section{Conclusions}
\label{conclus}
In conclusion, we have proposed and theoretically investigated an innovative highly sensitive magnetometer: the inductive superconducting quantum interference proximity transistor (L-SQUIPT). The L-SQUIPT promises enhanced performance with respect to  widespread SQUID and SQUIPT magnetometers. Indeed, an L-SQUIPTs made of conventional materials (such as aluminum and copper) would show a quantum limited intrinsic noise down to $\sim 8$ n$\Phi_0/\sqrt{\text{Hz}}$, both in current and voltage bias operations. Furthermore, the superconducting output probe allows to design two different dissipationless read-out schemes based on the variation of the Josephson inductance of the tunnel junction, such as inductive and dispersive read-out setups. In these configurations, the best flux fluctuation is $\average{\delta \Phi_L}\sim 4\;\mu\Phi_0$ for a bandwidth of 100 MHz. 

%%%%%%%%%%%%%%%%%%%%%%%%%%%%%%%%%%%%%%%%%%
\acknowledgments{We acknowledge A. Ronzani for useful discussions.
The authors acknowledge the European Research Council under Grant Agreement No. 899315 (TERASEC), and the EU’s Horizon 2020 research and innovation program under Grant Agreement No.  800923 (SUPERTED) and No. 964398 (SUPERGATE) for partial financial support.}

%\bibliographystyle{apsrev}
%\bibliography{L-Squipt_biblio}

\begin{thebibliography}{33}
\expandafter\ifx\csname natexlab\endcsname\relax\def\natexlab#1{#1}\fi
\expandafter\ifx\csname bibnamefont\endcsname\relax
  \def\bibnamefont#1{#1}\fi
\expandafter\ifx\csname bibfnamefont\endcsname\relax
  \def\bibfnamefont#1{#1}\fi
\expandafter\ifx\csname citenamefont\endcsname\relax
  \def\citenamefont#1{#1}\fi
\expandafter\ifx\csname url\endcsname\relax
  \def\url#1{\texttt{#1}}\fi
\expandafter\ifx\csname urlprefix\endcsname\relax\def\urlprefix{URL }\fi
\providecommand{\bibinfo}[2]{#2}
\providecommand{\eprint}[2][]{\url{#2}}

\bibitem[{\citenamefont{Cantor and Koelle}(2004)}]{SQUIDhand}
\bibinfo{author}{\bibfnamefont{R.}~\bibnamefont{Cantor}} \bibnamefont{and}
  \bibinfo{author}{\bibfnamefont{D.}~\bibnamefont{Koelle}},
  \emph{\bibinfo{title}{Practical DC SQUIDS: Configuration and Performance}}
  (\bibinfo{publisher}{John Wiley \& Sons, Ltd}, \bibinfo{year}{2004}),
  chap.~\bibinfo{chapter}{5}, pp. \bibinfo{pages}{171--217}, ISBN
  \bibinfo{isbn}{9783527603640}.

\bibitem[{\citenamefont{Jaklevic et~al.}(1964)\citenamefont{Jaklevic, Lambe,
  Silver, and Mercereau}}]{Jaklevic1964}
\bibinfo{author}{\bibfnamefont{R.~C.} \bibnamefont{Jaklevic}},
  \bibinfo{author}{\bibfnamefont{J.}~\bibnamefont{Lambe}},
  \bibinfo{author}{\bibfnamefont{A.~H.} \bibnamefont{Silver}},
  \bibnamefont{and} \bibinfo{author}{\bibfnamefont{J.~E.}
  \bibnamefont{Mercereau}}, \bibinfo{journal}{Phys. Rev. Lett.}
  \textbf{\bibinfo{volume}{12}}, \bibinfo{pages}{159} (\bibinfo{year}{1964}),
  \urlprefix\url{https://link.aps.org/doi/10.1103/PhysRevLett.12.159}.

\bibitem[{\citenamefont{Vasyukov et~al.}(2013)\citenamefont{Vasyukov, Anahory,
  Embon, Halbertal, Cuppens, Neeman, Finkler, Segev, Myasoedov, Rappaport
  et~al.}}]{Vasyukov2013}
\bibinfo{author}{\bibfnamefont{D.}~\bibnamefont{Vasyukov}},
  \bibinfo{author}{\bibfnamefont{Y.}~\bibnamefont{Anahory}},
  \bibinfo{author}{\bibfnamefont{L.}~\bibnamefont{Embon}},
  \bibinfo{author}{\bibfnamefont{D.}~\bibnamefont{Halbertal}},
  \bibinfo{author}{\bibfnamefont{J.}~\bibnamefont{Cuppens}},
  \bibinfo{author}{\bibfnamefont{L.}~\bibnamefont{Neeman}},
  \bibinfo{author}{\bibfnamefont{A.}~\bibnamefont{Finkler}},
  \bibinfo{author}{\bibfnamefont{Y.}~\bibnamefont{Segev}},
  \bibinfo{author}{\bibfnamefont{Y.}~\bibnamefont{Myasoedov}},
  \bibinfo{author}{\bibfnamefont{M.~L.} \bibnamefont{Rappaport}},
  \bibnamefont{et~al.}, \bibinfo{journal}{Nature Nanotech}
  \textbf{\bibinfo{volume}{8}}, \bibinfo{pages}{639} (\bibinfo{year}{2013}),
  \urlprefix\url{https://link.aps.org/doi/10.1103/PhysRevB.85.024527}.

\bibitem[{\citenamefont{Giazotto et~al.}(2010)\citenamefont{Giazotto, Peltonen,
  Meschke, and Pekola}}]{Giazotto2010}
\bibinfo{author}{\bibfnamefont{F.}~\bibnamefont{Giazotto}},
  \bibinfo{author}{\bibfnamefont{J.~T.} \bibnamefont{Peltonen}},
  \bibinfo{author}{\bibfnamefont{M.}~\bibnamefont{Meschke}}, \bibnamefont{and}
  \bibinfo{author}{\bibfnamefont{J.~P.} \bibnamefont{Pekola}},
  \bibinfo{journal}{Nature Physics} \textbf{\bibinfo{volume}{6}},
  \bibinfo{pages}{254} (\bibinfo{year}{2010}),
  \urlprefix\url{https://www.nature.com/articles/nphys1537}.

\bibitem[{\citenamefont{Meschke et~al.}(2011)\citenamefont{Meschke, Peltonen,
  Pekola, and Giazotto}}]{Meschke2011}
\bibinfo{author}{\bibfnamefont{M.}~\bibnamefont{Meschke}},
  \bibinfo{author}{\bibfnamefont{J.~T.} \bibnamefont{Peltonen}},
  \bibinfo{author}{\bibfnamefont{J.~P.} \bibnamefont{Pekola}},
  \bibnamefont{and} \bibinfo{author}{\bibfnamefont{F.}~\bibnamefont{Giazotto}},
  \bibinfo{journal}{Phys. Rev. B} \textbf{\bibinfo{volume}{84}},
  \bibinfo{pages}{214514} (\bibinfo{year}{2011}),
  \urlprefix\url{https://link.aps.org/doi/10.1103/PhysRevB.84.214514}.

\bibitem[{\citenamefont{Ronzani et~al.}(2014)\citenamefont{Ronzani, Altimiras,
  and Giazotto}}]{Ronzani2014}
\bibinfo{author}{\bibfnamefont{A.}~\bibnamefont{Ronzani}},
  \bibinfo{author}{\bibfnamefont{C.}~\bibnamefont{Altimiras}},
  \bibnamefont{and} \bibinfo{author}{\bibfnamefont{F.}~\bibnamefont{Giazotto}},
  \bibinfo{journal}{Phys. Rev. Applied} \textbf{\bibinfo{volume}{2}},
  \bibinfo{pages}{024005} (\bibinfo{year}{2014}),
  \urlprefix\url{https://link.aps.org/doi/10.1103/PhysRevApplied.2.024005}.

\bibitem[{\citenamefont{Jabdaraghi et~al.}(2014)\citenamefont{Jabdaraghi,
  Meschke, and Pekola}}]{Jabdaraghi2014}
\bibinfo{author}{\bibfnamefont{R.~N.} \bibnamefont{Jabdaraghi}},
  \bibinfo{author}{\bibfnamefont{M.}~\bibnamefont{Meschke}}, \bibnamefont{and}
  \bibinfo{author}{\bibfnamefont{J.~P.} \bibnamefont{Pekola}},
  \bibinfo{journal}{Applied Physics Letters} \textbf{\bibinfo{volume}{104}},
  \bibinfo{pages}{082601} (\bibinfo{year}{2014}),
  \eprint{https://doi.org/10.1063/1.4866584},
  \urlprefix\url{https://doi.org/10.1063/1.4866584}.

\bibitem[{\citenamefont{D'Ambrosio et~al.}(2015)\citenamefont{D'Ambrosio,
  Meissner, Blanc, Ronzani, and Giazotto}}]{Sophie2015}
\bibinfo{author}{\bibfnamefont{S.}~\bibnamefont{D'Ambrosio}},
  \bibinfo{author}{\bibfnamefont{M.}~\bibnamefont{Meissner}},
  \bibinfo{author}{\bibfnamefont{C.}~\bibnamefont{Blanc}},
  \bibinfo{author}{\bibfnamefont{A.}~\bibnamefont{Ronzani}}, \bibnamefont{and}
  \bibinfo{author}{\bibfnamefont{F.}~\bibnamefont{Giazotto}},
  \bibinfo{journal}{Applied Physics Letters} \textbf{\bibinfo{volume}{107}},
  \bibinfo{pages}{113110} (\bibinfo{year}{2015}),
  \eprint{https://doi.org/10.1063/1.4930934},
  \urlprefix\url{https://doi.org/10.1063/1.4930934}.

\bibitem[{\citenamefont{Jabdaraghi et~al.}(2016)\citenamefont{Jabdaraghi,
  Peltonen, Saira, and Pekola}}]{Jabdaraghi2016}
\bibinfo{author}{\bibfnamefont{R.~N.} \bibnamefont{Jabdaraghi}},
  \bibinfo{author}{\bibfnamefont{J.~T.} \bibnamefont{Peltonen}},
  \bibinfo{author}{\bibfnamefont{O.-P.} \bibnamefont{Saira}}, \bibnamefont{and}
  \bibinfo{author}{\bibfnamefont{J.~P.} \bibnamefont{Pekola}},
  \bibinfo{journal}{Applied Physics Letters} \textbf{\bibinfo{volume}{108}},
  \bibinfo{pages}{042604} (\bibinfo{year}{2016}),
  \eprint{https://doi.org/10.1063/1.4940979},
  \urlprefix\url{https://doi.org/10.1063/1.4940979}.

\bibitem[{\citenamefont{Jabdaraghi et~al.}(2017)\citenamefont{Jabdaraghi,
  Golubev, Pekola, and Peltonen}}]{Jabdaraghi2017}
\bibinfo{author}{\bibfnamefont{R.~N.} \bibnamefont{Jabdaraghi}},
  \bibinfo{author}{\bibfnamefont{D.~S.} \bibnamefont{Golubev}},
  \bibinfo{author}{\bibfnamefont{J.~P.} \bibnamefont{Pekola}},
  \bibnamefont{and} \bibinfo{author}{\bibfnamefont{J.~T.}
  \bibnamefont{Peltonen}}, \bibinfo{journal}{Scientific Reports}
  \textbf{\bibinfo{volume}{7}}, \bibinfo{pages}{8011} (\bibinfo{year}{2017}),
  \eprint{https://doi.org/10.1038/s41598-017-08710-7},
  \urlprefix\url{https://doi.org/10.1038/s41598-017-08710-7}.

\bibitem[{\citenamefont{Virtanen et~al.}(2016)\citenamefont{Virtanen, Ronzani,
  and Giazotto}}]{Virtanen2016}
\bibinfo{author}{\bibfnamefont{P.}~\bibnamefont{Virtanen}},
  \bibinfo{author}{\bibfnamefont{A.}~\bibnamefont{Ronzani}}, \bibnamefont{and}
  \bibinfo{author}{\bibfnamefont{F.}~\bibnamefont{Giazotto}},
  \bibinfo{journal}{Phys. Rev. Applied} \textbf{\bibinfo{volume}{6}},
  \bibinfo{pages}{054002} (\bibinfo{year}{2016}),
  \urlprefix\url{https://link.aps.org/doi/10.1103/PhysRevApplied.6.054002}.

\bibitem[{\citenamefont{Ronzani et~al.}(2017)\citenamefont{Ronzani, D'Ambrosio,
  Virtanen, Giazotto, and Altimiras}}]{Ronzani2017}
\bibinfo{author}{\bibfnamefont{A.}~\bibnamefont{Ronzani}},
  \bibinfo{author}{\bibfnamefont{S.}~\bibnamefont{D'Ambrosio}},
  \bibinfo{author}{\bibfnamefont{P.}~\bibnamefont{Virtanen}},
  \bibinfo{author}{\bibfnamefont{F.}~\bibnamefont{Giazotto}}, \bibnamefont{and}
  \bibinfo{author}{\bibfnamefont{C.}~\bibnamefont{Altimiras}},
  \bibinfo{journal}{Phys. Rev. B} \textbf{\bibinfo{volume}{96}},
  \bibinfo{pages}{214517} (\bibinfo{year}{2017}),
  \urlprefix\url{https://link.aps.org/doi/10.1103/PhysRevB.96.214517}.

\bibitem[{\citenamefont{Ligato et~al.}(2016)\citenamefont{Ligato, Marchegiani,
  Virtanen, Strambini, and Giazotto}}]{Ligato2017}
\bibinfo{author}{\bibfnamefont{N.}~\bibnamefont{Ligato}},
  \bibinfo{author}{\bibfnamefont{G.}~\bibnamefont{Marchegiani}},
  \bibinfo{author}{\bibfnamefont{P.}~\bibnamefont{Virtanen}},
  \bibinfo{author}{\bibfnamefont{E.}~\bibnamefont{Strambini}},
  \bibnamefont{and} \bibinfo{author}{\bibfnamefont{F.}~\bibnamefont{Giazotto}},
  \bibinfo{journal}{Scientific Reports} \textbf{\bibinfo{volume}{7}},
  \bibinfo{pages}{8810} (\bibinfo{year}{2016}),
  \urlprefix\url{https://www.nature.com/articles/s41598-017-09036-0}.

\bibitem[{\citenamefont{De~Gennes}(1999)}]{DeGennes}
\bibinfo{author}{\bibfnamefont{P.~G.} \bibnamefont{De~Gennes}},
  \emph{\bibinfo{title}{{Superconductivity of Metals and Alloys}}}, Advanced
  book classics (\bibinfo{publisher}{Perseus}, \bibinfo{address}{Cambridge,
  MA}, \bibinfo{year}{1999}),
  \urlprefix\url{https://cds.cern.ch/record/566105}.

\bibitem[{\citenamefont{Buzdin}(2005)}]{Buzdin2005}
\bibinfo{author}{\bibfnamefont{A.~I.} \bibnamefont{Buzdin}},
  \bibinfo{journal}{Rev. Mod. Phys.} \textbf{\bibinfo{volume}{77}},
  \bibinfo{pages}{935} (\bibinfo{year}{2005}),
  \urlprefix\url{https://link.aps.org/doi/10.1103/RevModPhys.77.935}.

\bibitem[{\citenamefont{Zhou et~al.}(1998)\citenamefont{Zhou, Charlat, Spivak,
  and Pannetier}}]{Zhou1998}
\bibinfo{author}{\bibfnamefont{F.}~\bibnamefont{Zhou}},
  \bibinfo{author}{\bibfnamefont{P.}~\bibnamefont{Charlat}},
  \bibinfo{author}{\bibfnamefont{B.}~\bibnamefont{Spivak}}, \bibnamefont{and}
  \bibinfo{author}{\bibfnamefont{B.}~\bibnamefont{Pannetier}},
  \bibinfo{journal}{Journal of Low Temperature Physics}
  \textbf{\bibinfo{volume}{110}}, \bibinfo{pages}{841} (\bibinfo{year}{1998}),
  \urlprefix\url{https://doi.org/10.1023/A:1022628927203}.

\bibitem[{\citenamefont{Giazotto and Taddei}(2011)}]{Giazotto2011}
\bibinfo{author}{\bibfnamefont{F.}~\bibnamefont{Giazotto}} \bibnamefont{and}
  \bibinfo{author}{\bibfnamefont{F.}~\bibnamefont{Taddei}},
  \bibinfo{journal}{Phys. Rev. B} \textbf{\bibinfo{volume}{84}},
  \bibinfo{pages}{214502} (\bibinfo{year}{2011}),
  \urlprefix\url{https://link.aps.org/doi/10.1103/PhysRevB.84.214502}.

\bibitem[{\citenamefont{Golubov et~al.}(2004)\citenamefont{Golubov, Kupriyanov,
  and Il'ichev}}]{Golubov2004}
\bibinfo{author}{\bibfnamefont{A.~A.} \bibnamefont{Golubov}},
  \bibinfo{author}{\bibfnamefont{M.~Y.} \bibnamefont{Kupriyanov}},
  \bibnamefont{and} \bibinfo{author}{\bibfnamefont{E.}~\bibnamefont{Il'ichev}},
  \bibinfo{journal}{Rev. Mod. Phys.} \textbf{\bibinfo{volume}{76}},
  \bibinfo{pages}{411} (\bibinfo{year}{2004}),
  \urlprefix\url{https://link.aps.org/doi/10.1103/RevModPhys.76.411}.

\bibitem[{\citenamefont{Likharev}(1979)}]{Likharev1979}
\bibinfo{author}{\bibfnamefont{K.~K.} \bibnamefont{Likharev}},
  \bibinfo{journal}{Rev. Mod. Phys.} \textbf{\bibinfo{volume}{51}},
  \bibinfo{pages}{101} (\bibinfo{year}{1979}),
  \urlprefix\url{https://doi.org/10.1103/RevModPhys.51.101}.

\bibitem[{\citenamefont{le~Sueur et~al.}(2008)\citenamefont{le~Sueur, Joyez,
  Pothier, Urbina, and Esteve}}]{leSueur2008}
\bibinfo{author}{\bibfnamefont{H.}~\bibnamefont{le~Sueur}},
  \bibinfo{author}{\bibfnamefont{P.}~\bibnamefont{Joyez}},
  \bibinfo{author}{\bibfnamefont{H.}~\bibnamefont{Pothier}},
  \bibinfo{author}{\bibfnamefont{C.}~\bibnamefont{Urbina}}, \bibnamefont{and}
  \bibinfo{author}{\bibfnamefont{D.}~\bibnamefont{Esteve}},
  \bibinfo{journal}{Phys. Rev. Lett.} \textbf{\bibinfo{volume}{100}},
  \bibinfo{pages}{197002} (\bibinfo{year}{2008}),
  \urlprefix\url{https://link.aps.org/doi/10.1103/PhysRevLett.100.197002}.

\bibitem[{\citenamefont{Heikkil\"a et~al.}(2002)\citenamefont{Heikkil\"a,
  S\"arkk\"a, and Wilhelm}}]{Heikkila2002}
\bibinfo{author}{\bibfnamefont{T.~T.} \bibnamefont{Heikkil\"a}},
  \bibinfo{author}{\bibfnamefont{J.}~\bibnamefont{S\"arkk\"a}},
  \bibnamefont{and} \bibinfo{author}{\bibfnamefont{F.~K.}
  \bibnamefont{Wilhelm}}, \bibinfo{journal}{Phys. Rev. B}
  \textbf{\bibinfo{volume}{66}}, \bibinfo{pages}{184513}
  (\bibinfo{year}{2002}),
  \urlprefix\url{https://link.aps.org/doi/10.1103/PhysRevB.66.184513}.

\bibitem[{\citenamefont{Ligato et~al.}(2021)\citenamefont{Ligato, Strambini,
  Paolucci, and Giazotto}}]{ligato2020}
\bibinfo{author}{\bibfnamefont{N.}~\bibnamefont{Ligato}},
  \bibinfo{author}{\bibfnamefont{E.}~\bibnamefont{Strambini}},
  \bibinfo{author}{\bibfnamefont{F.}~\bibnamefont{Paolucci}}, \bibnamefont{and}
  \bibinfo{author}{\bibfnamefont{F.}~\bibnamefont{Giazotto}},
  \bibinfo{journal}{Nature Communications} \textbf{\bibinfo{volume}{12}},
  \bibinfo{pages}{5200} (\bibinfo{year}{2021}),
  \urlprefix\url{https://doi.org/10.1038/s41467-021-25209-y}.

\bibitem[{\citenamefont{Polini et~al.}(2022)\citenamefont{Polini, Giazotto,
  Fong, Pop, Schuck, Boccali, Signorelli, D'Elia, Hadfield, Giovannetti
  et~al.}}]{polini2022materials}
\bibinfo{author}{\bibfnamefont{M.}~\bibnamefont{Polini}},
  \bibinfo{author}{\bibfnamefont{F.}~\bibnamefont{Giazotto}},
  \bibinfo{author}{\bibfnamefont{K.~C.} \bibnamefont{Fong}},
  \bibinfo{author}{\bibfnamefont{I.~M.} \bibnamefont{Pop}},
  \bibinfo{author}{\bibfnamefont{C.}~\bibnamefont{Schuck}},
  \bibinfo{author}{\bibfnamefont{T.}~\bibnamefont{Boccali}},
  \bibinfo{author}{\bibfnamefont{G.}~\bibnamefont{Signorelli}},
  \bibinfo{author}{\bibfnamefont{M.}~\bibnamefont{D'Elia}},
  \bibinfo{author}{\bibfnamefont{R.~H.} \bibnamefont{Hadfield}},
  \bibinfo{author}{\bibfnamefont{V.}~\bibnamefont{Giovannetti}},
  \bibnamefont{et~al.}, \bibinfo{journal}{arXiv preprint arXiv:2201.09260}
  (\bibinfo{year}{2022}).

\bibitem[{\citenamefont{Kulik and Omel'yanchuk}(1975)}]{Kulik1975}
\bibinfo{author}{\bibfnamefont{I.~O.} \bibnamefont{Kulik}} \bibnamefont{and}
  \bibinfo{author}{\bibfnamefont{A.~N.} \bibnamefont{Omel'yanchuk}},
  \bibinfo{journal}{JEPT Lett.} \textbf{\bibinfo{volume}{21}},
  \bibinfo{pages}{96} (\bibinfo{year}{1975}).

\bibitem[{\citenamefont{Artemenko et~al.}(1979)\citenamefont{Artemenko, Volkov,
  and Zaitsev}}]{Artemenko1979}
\bibinfo{author}{\bibfnamefont{S.~N.} \bibnamefont{Artemenko}},
  \bibinfo{author}{\bibfnamefont{A.~F.} \bibnamefont{Volkov}},
  \bibnamefont{and} \bibinfo{author}{\bibfnamefont{A.~V.}
  \bibnamefont{Zaitsev}}, \bibinfo{journal}{Sov. Phys. JEPT}
  \textbf{\bibinfo{volume}{49}}, \bibinfo{pages}{924} (\bibinfo{year}{1979}).

\bibitem[{\citenamefont{Dynes et~al.}(1984)\citenamefont{Dynes, Garno, Hertel,
  and Orlando}}]{Dynes1984}
\bibinfo{author}{\bibfnamefont{R.~C.} \bibnamefont{Dynes}},
  \bibinfo{author}{\bibfnamefont{J.~P.} \bibnamefont{Garno}},
  \bibinfo{author}{\bibfnamefont{G.~B.} \bibnamefont{Hertel}},
  \bibnamefont{and} \bibinfo{author}{\bibfnamefont{T.~P.}
  \bibnamefont{Orlando}}, \bibinfo{journal}{Phys. Rev. Lett.}
  \textbf{\bibinfo{volume}{53}}, \bibinfo{pages}{2437} (\bibinfo{year}{1984}),
  \urlprefix\url{https://link.aps.org/doi/10.1103/PhysRevLett.53.2437}.

\bibitem[{\citenamefont{Tinkham}(2012)}]{tinkham2012introduction}
\bibinfo{author}{\bibfnamefont{M.}~\bibnamefont{Tinkham}},
  \emph{\bibinfo{title}{Introduction to superconductivity}}
  (\bibinfo{publisher}{Courier Dover Publications}, \bibinfo{year}{2012}).

\bibitem[{\citenamefont{Kirtley}(2010)}]{Kirtley2010}
\bibinfo{author}{\bibfnamefont{J.~R.} \bibnamefont{Kirtley}},
  \bibinfo{journal}{Reports on Progress in Physics}
  \textbf{\bibinfo{volume}{73}}, \bibinfo{pages}{126501}
  (\bibinfo{year}{2010}),
  \urlprefix\url{https://doi.org/10.1088/0034-4885/73/12/126501}.

\bibitem[{\citenamefont{Ambegaokar and Baratoff}(1963)}]{Ambegaokar1963}
\bibinfo{author}{\bibfnamefont{V.}~\bibnamefont{Ambegaokar}} \bibnamefont{and}
  \bibinfo{author}{\bibfnamefont{A.}~\bibnamefont{Baratoff}},
  \bibinfo{journal}{Phys. Rev. Lett.} \textbf{\bibinfo{volume}{10}},
  \bibinfo{pages}{486} (\bibinfo{year}{1963}),
  \urlprefix\url{https://link.aps.org/doi/10.1103/PhysRevLett.10.486}.

\bibitem[{\citenamefont{Giazotto et~al.}(2008)\citenamefont{Giazotto,
  Heikkilä, Pepe, Helistö, Luukanen, and Pekola}}]{Giazotto2008}
\bibinfo{author}{\bibfnamefont{F.}~\bibnamefont{Giazotto}},
  \bibinfo{author}{\bibfnamefont{T.~T.} \bibnamefont{Heikkilä}},
  \bibinfo{author}{\bibfnamefont{G.~P.} \bibnamefont{Pepe}},
  \bibinfo{author}{\bibfnamefont{P.}~\bibnamefont{Helistö}},
  \bibinfo{author}{\bibfnamefont{A.}~\bibnamefont{Luukanen}}, \bibnamefont{and}
  \bibinfo{author}{\bibfnamefont{J.~P.} \bibnamefont{Pekola}},
  \bibinfo{journal}{Applied Physics Letters} \textbf{\bibinfo{volume}{92}},
  \bibinfo{pages}{162507} (\bibinfo{year}{2008}),
  \eprint{https://doi.org/10.1063/1.2908922},
  \urlprefix\url{https://doi.org/10.1063/1.2908922}.

\bibitem[{\citenamefont{Barone and Paterno}(1982)}]{BaronePaterno1982}
\bibinfo{author}{\bibfnamefont{A.}~\bibnamefont{Barone}} \bibnamefont{and}
  \bibinfo{author}{\bibfnamefont{G.}~\bibnamefont{Paterno}},
  \emph{\bibinfo{title}{Physics and applications of Josephson effect}}
  (\bibinfo{publisher}{Wiley New York}, \bibinfo{year}{1982}), ISBN
  \bibinfo{isbn}{0471014699,9780471014690}.

\bibitem[{\citenamefont{Guarcello et~al.}(2018)\citenamefont{Guarcello,
  Solinas, Braggio, Di~Ventra, and Giazotto}}]{guarcello2018}
\bibinfo{author}{\bibfnamefont{C.}~\bibnamefont{Guarcello}},
  \bibinfo{author}{\bibfnamefont{P.}~\bibnamefont{Solinas}},
  \bibinfo{author}{\bibfnamefont{A.}~\bibnamefont{Braggio}},
  \bibinfo{author}{\bibfnamefont{M.}~\bibnamefont{Di~Ventra}},
  \bibnamefont{and} \bibinfo{author}{\bibfnamefont{F.}~\bibnamefont{Giazotto}},
  \bibinfo{journal}{Phys. Rev. Applied} \textbf{\bibinfo{volume}{9}},
  \bibinfo{pages}{014021} (\bibinfo{year}{2018}),
  \urlprefix\url{https://link.aps.org/doi/10.1103/PhysRevApplied.9.014021}.

\bibitem[{\citenamefont{Solinas et~al.}(2012)\citenamefont{Solinas,
  M\"ott\"onen, Salmilehto, and Pekola}}]{solinas2012}
\bibinfo{author}{\bibfnamefont{P.}~\bibnamefont{Solinas}},
  \bibinfo{author}{\bibfnamefont{M.}~\bibnamefont{M\"ott\"onen}},
  \bibinfo{author}{\bibfnamefont{J.}~\bibnamefont{Salmilehto}},
  \bibnamefont{and} \bibinfo{author}{\bibfnamefont{J.~P.}
  \bibnamefont{Pekola}}, \bibinfo{journal}{Phys. Rev. B}
  \textbf{\bibinfo{volume}{85}}, \bibinfo{pages}{024527}
  (\bibinfo{year}{2012}),
  \urlprefix\url{https://link.aps.org/doi/10.1103/PhysRevB.85.024527}.

\end{thebibliography}

\end{document}